\documentclass[sigconf, nonacm]{acmart}
\usepackage{tabularx}
\AtBeginDocument{%
  }

\begin{document}

\title{Closing the Loop: A Software Framework for AI to Support Business Decision Making}

\author{Jeffrey Wong}
\affiliation{%
  \institution{Airbnb}
  \city{San Francisco, California}
  \country{USA}}
\email{jeffrey.wong@airbnb.com}

\author{Antoine Creux}
\affiliation{%
  \institution{Airbnb}
  \city{San Francisco, California}
  \country{USA}}
\email{antoine.creux@airbnb.com}

\renewcommand{\shortauthors}{Wong and Creux}

\begin{abstract}
Create an idea, prototype it, evaluate if users like it, then learn. It is the circle of business. If AI can operate in all parts of the circle, it will enable rapid iteration and learning speeds for businesses. Experiment platforms that deploy experiments to evaluate return on investment for businesses are abundant, but systems that help businesses learn personalization, mechanisms, and what to ideate next, are rare. Among technologies that do exist, they cannot be well orchestrated in a single software interface that can be safely and efficiently leveraged by an AI agent. These challenges make it difficult to teach an AI agent how to learn within a robust experimentation framework, and difficult for an AI agent to operate and iterate for the business.

We offer a two part solution: one half that is rooted in mathematical reductions to contain complexity, and one half that is rooted in software design to optimize for orchestration, software safety, and multiplicity. Our solution, a software framework, moves beyond the simple treatment effect computed as a difference in means. To create a better understanding of a business and its customers, we enrich causal analysis with heterogeneous effects, policy algorithms, mediation analysis, and forecasts of effects. To have an AI complete the iteration cycle faster, we further enrich the analysis with variance reduction and anytime valid inference. The enrichments are made compatible across different types of experiments, and are presented in a single software interface that is usable in an AI agent.

We evaluate the approach on various objectives in experiment analysis, and show that the framework improves code correctness, reduces lines of code, and is more performant than a baseline analysis constructed by a vanilla agent.

\end{abstract}




\maketitle

\section{Introduction}

AI's value proposition for businesses is only half complete. It is helping to ideate at spectacular speed. However, we also need it to evaluate those ideas in rigorous experiments. Firms will then be able to identify and eliminate bad ideas, scale good ideas and return to ideate on new ones. In a study of over 35,000 technology startups, those that used AB tests to evaluate ideas were able to introduce new products faster, were able to identify positive tail outcomes, and had more page views \cite{koning2022experimentation}. The program of repeated experimentation led firms to succeed. Ideation, and evaluation, must exist in an iterative loop.

A simple flywheel spanning ideation and evaluation has been discussed in the past 15 years, and there is an opportunity to use AI to greatly magnify the previous success. The flywheel is focused on enabling people with data: when a business wants to make a change, teams need to do quantitative research on the addressable market, potential returns on investments, opportunity costs, and competition. AI is helping businesses conduct that prerequisite research. Second, new ideas need to be implemented and incremental versions need to be tested online in a controlled experiment. Companies such as Eppo and LaunchDarkly are offering services to integrate their platform for online experiments with various products. Furthermore, AI agents such as Positron and Colab allow us to analyze data with AI. With low cost experimentation programs, businesses can quantify returns and use them as inputs for decision making. This can be done using AI interfaces today.

The value proposition is still not complete. Despite experimentation becoming a driver for whether or not a business releases a feature, it is still difficult to understand how to iterate on a feature \cite{gupta2019top, felin2017theory}. Innovating from the experiment report to the next phase of ideation is still a manual and time consuming process. Businesses need to know why their feature changes are yielding positive or negative returns. They need to synthesize a coherent understanding of the many metrics that increase and decrease. They need to understand the causal mechanism that relates the feature change, user behavior, and the metrics. They need to understand how their feature changes motivate different audience segments. Interpretable and actionable results are key. All of this needs to be analyzed in a statistically robust way that respects the experiment and causal inference. While AI is assisting in market research, creating prototypes and experiments, and analyzing for ROI, the flywheel step from reporting to insight is not well covered by AI tools.

Our paper focuses on this aspect of business iteration, and what foundational software is needed for AI to close this gap. Estimating the treatment effect in an experiment using the difference in two averages is not enough. At the same time, research has made advancements in the theory and practice of experimentation, specifically in:
\begin{enumerate}
    \item Identifying the treatment effect faster, which will tighten the loop.
    \item Estimating heterogeneous effects, which will allow businesses to understand market segments better.
    \item Identifying causal mechanisms, which will help businesses to understand which customer interactions lead to positive or negative ROI.
\end{enumerate}
While software exists to implement this research, there are many practical challenges. Implementations are scattered across multiple software libraries, with different data structures and interfaces, making it difficult for an AI agent to \textbf{orchestrate} across them correctly. Next, there is doubt that AI can reason with cause and effect. The lack of causal inference principles creates risk when businesses analyze experiments: the AI may generate code that uses randomization from the data generating process incorrectly, and return correlational estimates instead of causal estimates. The sign of the effect may be incorrect. The magnitude of the effect may be exaggerated. If the mechanism for the effect is wrong, it can mislead business strategy for the next iteration. \textbf{Software safety} is critical in order to match the robustness and trust inherent in controlled experiments. Finally, most software solutions are not optimized for \textbf{multiplicity}. All of these challenges warrant a new solution that is uniquely rooted in the mathematical frameworks of causal inference, software design, and human-AI interaction.

Our solution is a unique mixture of mathematical reductions, and software framework. The framework unifies many disciplines within causal inference into a single library and solves for orchestration, software safety, and multiplicity. The mathematical reductions look for commonalities across statistical models used for online experiments, and reduces them to a single interface. Support for models spans across regression, instrumental variables, heterogeneous treatment effects, anytime valid inference, policy algorithms, and mechanisms. The result is a software framework that can be invoked by an AI agent to produce a fast prompt-insight loop when analyzing experiments.  

\section{Related Works}

Our work synthesizes across data driven organizational leadership, causal inference, and software design. There are many studies across entrepreneurship and leadership that highlight the importance of agility and the cycle of ideation and evaluation. In \cite{reis2011lean}, the author recommends business models adopt the build-measure-learn paradigm. They emphasize that treating business as a scientific practice leads firms to have strong evidence as to what their customers want, how to build for them, and when to pivot. Later, a controlled experiment conducted on startups provided rigorous evidence that business leaders who were trained in scientific decision making were indeed able to recognize positive opportunities, and terminate negative ones. Their data reported large effects showing an increase in revenue \cite{camuffo2020scientific}. Felin and Zenger write many articles on a different perspective of business decision making, where experiments are an intentional part of a grand theory of the business. While they agree that a scientific approach and experiments are useful, they advocate that each experiment culminate in a theory of the business, which in turn informs the next experiment \cite{felin2017theory}. Underpinning this cycle is a robust learning of mechanisms that inform how the business should operate in its environment. Learning must move beyond whether an idea has positive returns or not.

The above works establish the importance of experiment analysis to firms, and the need for readily available analytical software. As we scope our work to the evaluation of ideas, especially in the domain of online experiments, we look to literature that describes how to create interpretable and actionable insights that are also trustworthy. Starting with \cite{gupta2019top, kohavi2022ab} the authors cite low statistical power, and a strong prior that most treatments do not create material change, as the reason why experiments have a high false positive risk. The ways to mitigate are to develop sensitive metrics and use models that can increase statistical power such as \cite{deng2013improving}. Risk compounds when companies peak, as described in \cite{johari2017peeking}. False positive risks challenge the trustworthiness of experiments and need to be addressed in each experiment analysis. Next, experiment insights can become actionable through personalization. \cite{sverdrup2025estimating, tu2021personalized} describes the importance of understanding heterogeneous effects as a means to personalize interventions. In the case of Netflix's personalization engine \cite{gomez2015netflix}, relating heterogeneity and personalization can be valued at over \$1 billion annually. 
Finally, studying mechanisms can make experiments interpretable. When a positive or negative outcome is detected, we want to attribute the result to a sequence of events that led to the outcome. This causal mediation analysis is described in  \cite{imai2010identification}. 
Understanding how a new product change effects user behavior and ultimately metrics is a key part of the AI that businesses need to make decisions on how to iterate a product.

It is not enough to study the mathematical framing of causal inference models. Here, we review what makes software malleable to a user's curiosity and state of flow. First, we start with OpenAI and Langchain's documentation around making effective AI agents \cite{openai_function_calling}. Both suggest developers to have few, but comprehensive functions that are tightly scoped. They recommend the principle of least surprise. Finally, they should be designed for iterative and human-in-the-loop feedback. Other than these principles, we draw inspiration from \cite{wickham_tidy_design_principles}, a set of software design principles governing the tidyverse. Such principles have enabled the tidyverse community to do exploratory data analysis (EDA) effectively. EDA has many parallels to interacting with conversational interfaces, such as designing software for curious exploration, and the ability to compose and exchange functions one at a time. While the community of tool developers for AI agents, and the community for EDA, have design principles, we have not found a comprehensive set of software primitives that are aligned to the mathematical frameworks of causal inference. Our work will specialize in the topic of causal analysis, where we leverage specific forms of online, controlled experiments to produce an effective software framework for AI Agents and humans.

\section{Motivating Example}

We will reference the following motivating example throughout the paper. Suppose an online store is considering changes to its ranking system. It believes that increasing the diversity of products shown will increase their sales. It can use AI to conduct market research and understand market segments where diversity will be relevant, and when it is not. For example, an AI can reply that holiday shoppers may be interested in such a diversity change, whereas return shoppers may not be. The change in the ranking system is implemented and is tested online in a controlled experiment.

The business is surprised to find that the effect on sales is negative. Many theories exist to explain why:

\begin{enumerate}
\item An increase in diversity makes it harder to find the product that the user was looking for, causing sales to decrease.
\item An increase in diversity creates decision fatigue, causing sales to decrease.
\item An increase in diversity is good, but the implementation was poor. Sales decrease because the website became slower. However, no decrease should be observed among shoppers that were on a desktop computer.
\end{enumerate}

Though an AI interface can implement an experiment in software and can report the negative return, the business needs to conduct a manual exercise to understand the causal mechanism for how an increase in diversity causes sales to decrease. To avoid an expensive and time consuming exercise, we wish that AI can deploy causal discovery algorithms to understand the mechanism. Afterwards, the business wants to understand opportunities for what to do next. They want to query an AI agent to understand whether there were other market segments, other than shoppers on desktop computers, that benefited from the increase in diversity. They also want to query if the poor implementation and website performance were resolved, would the diversity in the search cause an increase in sales. To provide an AI agent such analytical capabilities, we must convert the mathematical approach to data analysis for experiments into a software framework to interface with.

\section{Mathematical and Software Frameworks}
\label{objectives}

Our design for a software framework will unify many mathematical concepts of causal inference into a composable framework that is suitable for human-AI interaction.  The framework will enable the user to communicate chains of thoughts that are safely translated into numerical computation on experimental data. Collectively, it enables a fast prompt-insight loop that is statistically robust, and helps companies iterate through ideation and evaluation. We scope the framework for businesses that run online experiments, where the framework and computational model will achieve the following objectives:

\begin{enumerate}
    \item Create structure for AI to understand the different types of experiments that can be implemented, for example controlled and randomized experiments, cluster experiments \cite{donner2000design}, and encouragement designs \cite{hirano2000assessing}. The AI will understand how an effect is identified \cite{imbens1994identification}. \label{goal1}
    \item Apply appropriate statistical models based on the design of the experiment. \label{goal2}
    \item Use experimental data to estimate anytime valid treatment effects \cite{lindon2022anytime} with great precision \cite{deng2013improving}. \label{goal3}
    \item Iterate over a set of segments to estimate heterogeneous treatment effects and build a policy to determine which shoppers should receive the new treatments \cite{Wager03072018}. \label{goal4}
    \item Understand mechanisms for how the business change resulted in a change in metrics \cite{imai2010identification}. \label{goal5}
    \item Project the effect under different future distributions of shoppers. \label{goal6}
\end{enumerate}

\subsection{Designing for Real World Challenges}

To create such a framework, we will need to overcome challenges in orchestration, software safety, and multiplicity. Solutions to these challenges are crucial to make the AI agent viable, ensure statistically robust replies to chat-like prompts, and to have a clear and computationally efficient model. 
Our unique contribution to this field is in the conversion of existing mathematical solutions into a scalable and reliable software system.

Software implementations for the various mathematical frameworks of causal inference exist today. However, they are scattered across many different libraries with different interfaces, including dowhy 
\cite{sharma2020dowhy}, 
statsmodels 
\cite{seabold2010statsmodels}, 
causalML 
\cite{causalml_software}, 
econML 
\cite{econml}, avlm, grf \cite{Wager03072018} and more, making it difficult to connect inputs and outputs of software libraries, and for a user to prompt an AI to invoke them. Exchanging data and intermediate models across the libraries can be prohibitive, for example no interface exists to hypothesize a causal graph, estimate an anytime valid treatment effect from avlm, then construct an online policy using grf. The orchestration complexity mirrors those when choosing a machine learning architecture. Inspired by scikit-learn \cite{buitinck2013apidesignmachinelearning}, the tidyverse \cite{Wickham2019}, and tidymodels \cite{kuhn2022tidymodels}, our software framework contains a composable API that allows the user to customize their analysis and orchestrate across many disciplines in causal inference. Centering the framework around composable primitives allows the AI agent to translate the user chain of thoughts into a computational model. As the user evolves their chain of thought, the framework built on composable primitives makes it possible to change input data, models, and training and inference instructions one by one, with minimal error. 

The path to designing a causally motivated AI is nonlinear, especially in business environments. Above, we discussed the need to create a framework to orchestrate software implementations of the mathematical frameworks for experiment analysis. This is needed to make the AI agent viable. Now, we elaborate on the need for a software framework to contain business complexity. This is needed to make the AI agent practical.

Experiment analyses contain many forks. There are many different experimental designs that each create unique data generating processes, many different models to choose from that analyze experiment data differently, many different intermediate and outcome metrics to evaluate whether a business change is good for customers, and many causal graphs that explain how the business change induces a change in the metrics. We refer to this series of forks as \textbf{multiplicity}. Each fork can alter the conclusion of the experiment and therefore how the business should iterate. In order to construct an enterprise-grade AI agent, we must innovate new software structure to contain the complexity.

Multiplicity manifests not only in computation with many nested loops, it is a challenge for clarity of thought and usability. The user must remain focused on the business decision and the data, they should not be distracted by needing to manage forks. To resolve multiplicity, the design of our framework employs multiple \textbf{reduction} strategies, and showcases the unique manipulation of mathematical frameworks to produce better software. Throughout the implementation of our models, we minimize the number of classes that the user must interact with. Instead of implementing separate classes and functions for analyzing controlled experiments, cluster randomized experiments, and adaptive experiments, we reduce the underlying differences between the experiment designs and claim they are all special cases of a single experiment interface, eliminating one dimension of multiplicity. We repeat this reduction strategy for model training, and inference as well, where we claim many commonly used causal models are special cases of one model. Additionally, we reduce unnecessary numerical computation when training models by vectorizing computation across multiple arms and segments. This reduction in multiplicity is described in detail in the next section, and helps to resolve compute efficiency and clarity of thought when interacting with an AI agent for causal analysis.

\subsection{Promptability, Composability, and Human-Machine Interaction}

Creating an effective causal AI agent requires both an implementation of the mathematical algorithms for causal inference, as well as a solution for navigating the multiple layers of multiplicity. The solution must be simple to use in a prompt. To achieve this, we need to unify concepts across multiple disciplines that practice experimentation, then identify reusable primitives and functions that are common in all analyses. After this, we focus on composability, a software design pattern where a user is able to chain together functions as if they were chaining words into a sentence. A composable software framework benefits the human-computer interaction, where the experimenter can prompt an AI to analyze data, converse with it over multiple prompts, and extend the prompt, all while the machine has a systematic way to compose the appropriate functions into code to analyze data. The machine's ability to follow the experimenter as they navigate layers of multiplicity is crucial for developing an effective AI.

\section{The Software Framework}

\subsection{Randomization}

Business ideas that are tested online make use of randomization, whether through an AB test, a cluster randomized test, an encouragement design, or an adaptive test. We limit the scope of our framework to these families of controlled randomization, and avoid the complexities of observational methods. To develop a comprehensive framework, we need a data structure to inform an AI agent how randomization was deployed, which will later inform how a treatment effect is \textbf{identified} \cite{angrist2009mostly}, and how the statistical model must be fit. Our framework starts here.

There are five classes of variables needed to inform an AI agent how an effect is identified. This anchors all subsequent modeling.

\begin{enumerate}
    \item The intent to treat assignment, $A$. This describes what treatment this unit was randomly assigned.
    \item The treatment status, $T$. This describes whether the unit has successfully received the treatment.
    \item The propensity to be treated, $W$ \cite{rosenbaum1983central}. This describes the probability that the unit will be treated.
    \item The instrument, $Z$. This describes an instrumental variable (\cite{angrist1996identification}) that is used to project out nonrandom components of the treatment. In many business scenarios the instrument and the intent to treat assignment are the same, so $Z = A$.
    \item The clusters, $C$. This describes groups of units that are randomized together.
\end{enumerate}
Across these five variables, we can describe the randomization strategy of an AB test, a cluster randomized test, an encouragement design, and an adaptive test. The lightweight ExperimentData data structure shown below is the root of all analyses.

\begin{verbatim}
experimental_data = ExperimentData(
  intent_to_treat = ...,
  treated = ...,
  propensity = ...,
  instrument = ...,
  cluster_by = ...
)
\end{verbatim}

Our first reduction claims that the above tests can be reduced to a single interface using the variables above. The figure below describes the four experiments in three dimensions.
\begin{table}[h]
\centering
\small
\begin{tabularx}{\columnwidth}{l|XXX} 
\hline
 & Compliance & Covariance structure & Treatment Propensity \\ \hline
AB & Fully compliant & Diagonal & Constant \\
Encouragement & \textbf{Noncompliant} & Diagonal & Constant \\
Cluster & Fully compliant & \textbf{Block Diagonal} & Constant \\
Adaptive & Fully compliant & Diagonal & \textbf{Variable} \\ \hline
\end{tabularx}
\end{table}
The uniqueness of an encouragement design is that the users are encouraged to take the treatment, but are not forced to. In this environment, there is one sided compliance: users who are held back do not receive the treatment, and some users who are encouraged to take the treatment comply.
Next, regardless of the encouragement, groups of users in the data, or groups of observations, can have within-subject correlation. Finally, the adaptive test describes an environment where the propensity to be treated is a function of covariates, instead of a constant. We note that a test can exhibit any combination of these characteristics, such as an an encouragement design that is cluster randomized with adaptive targeting. Using this reduction we build the mathematical differences as primitives for our framework to satisfy objective \ref{goal1}, then make use of them in the training step below.

\subsection{Effect Estimation}
Next, we build models to estimate effects, typically the average treatment effect. Without structure, the four experiments would be analyzed with four models. AB tests would be analyzed using ordinary least squares (OLS), encouragement designs would be analyzed using two stage least squares (2SLS), cluster experiments would be analyzed using a variant of OLS, and adaptive tests would be analyzed using propensity weights. To further reduce the complexity of 16 possible API calls, we now invoke another reduction. We collapse many popular models into a single treatment effect model that is suitable for all four test types. The data stored in ExperimentData will allow us to invoke the appropriate model for the randomization structure, achieving objective \ref{goal2}. Call the outcome variable $y$, then we relate

\begin{align}
    y &= T \beta_1 + \varepsilon. & \text{The second stage.}\\
    T &= Z \pi_1 + \nu. & \text{The first stage.}
\end{align}
The distribution of $\varepsilon$ is assumed to be gaussian, with mean zero, and block diagonal covariances according to the clusters, C.

In this two stage least squares model (2SLS), we reduce the multiplicity around choosing a model to a single yet comprehensive interface for the AI agent. When starting with an AB test with full compliance, $Z = A = T$, the two stage least squares model has $\pi_1 = 1$ and the second stage reduces to the classic ordinary least squares (OLS) model used to analyze AB tests. In the case of an encouragement design, the compliance is variable and the compliance rate is reflected in $\pi_1$. $\beta_1$ then becomes the estimate of the local average effect for the encouragement design. For clustered experiments, the estimation of the parameters is the same as estimating the AB test, but the structure of the covariances is different. Using the generic sandwich estimator \cite{white1980heteroskedasticity}, the covariance of the estimates is


\begin{align}
\varepsilon = y - T\beta. & & \hat{T}_c = (P_Z T)_c. & & \hat{T}_i &= (P_Z T)_i.
\end{align}

\begin{align}
Cov(\hat{\beta}_\text{clustered}) &= (T^T P_Z T)^{-1} \bigl(\sum_{c=1}^C \hat{T}_c^T \varepsilon_c \varepsilon_c^T \hat{T}_c \bigr) (T^T P_Z T)^{-1}. \\
Cov(\hat{\beta}_\text{OLS}) &= (T^T P_Z T)^{-1} \bigl(\sum_{i=1}^n \hat{T}_i^T \hat{T}_i \varepsilon_i^2 \bigr) (T^T P_Z T)^{-1}.
\end{align}

Under the case that each cluster has only one observation, and therefore reflects independent data, the formula for clustered covariances reduces to robust covariances for OLS. By allowing the covariance matrix for $\varepsilon$ to be block diagonal, instead of strictly diagonal, we can accommodate experiments that need to be analyzed with OLS and experiments that need clustered standard errors. Finally, an adaptive test collects data with variable treatment propensity. These can be passed to a 2SLS model as weights. 
Collectively, we have reduced the different experiment types and models to one interface, which operate on the same linear model and ExperimentData structure. This minimizes the risk that the AI agent uses the wrong model for a given experiment type, and reduces complexity from multiplicity.

Achieving the first two objectives allows an AI agent to understand randomization structure and fit the naive form of the treatment effect analysis. To make iteration speed fast, the analysis needs to permit the business to make conclusions as fast as possible. We invoke a third reduction on hypothesis testing. Most experiments run for a fixed horizon; conclusions cannot be drawn until the end of the test when sufficient statistical power has been accumulated. While this is an important practice to minimize false positives, this slows the iteration speed. Peeking at results and drawing daily, intermediate conclusions is a natural practice, though it creates an exponential probability of encountering a false positive. Anytime-valid inference allows the business to peek at results safely by forming a $(1 - \alpha)$\% confidence sequence. The strategy is a modification at inference time. One can reuse the treatment effect output from OLS, and its standard error, to compute an anytime valid p value for hypothesis testing \cite{lindon2022anytime}. Therefore, the AI agent is able to use peeking in a safe way that is compatible with our previous reduction strategies and is therefore compatible across all types of tests with minimal multiplicity.

To further speed up iteration, we minimize the variance of the estimator, which will also increase statistical power. This can be done by adding exogeneous covariates, $X$, to linear models so that 

\begin{align}
    y = T \beta_1 + f(X)\beta_2 + (T \cdot f(X))\beta_3 + \varepsilon. \\
    [T, f(X), (T \cdot f(X))] = Z \pi_1 + f(X)\pi_2 + (Z \cdot f(X))\pi_3 + \nu.
\end{align}
Adding arbitrary covariates can be done safely as long as the AI agent understands that they were measured prior to the unit receiving treatment. Therefore, it can be applied to all types of experiments, whether there is an instrument, a propensity weight, cluster randomization, or anytime-valid inference. The increase in speed achieves objective \ref{goal3}.

To train a model, our framework leverages the structure of the randomization specified in ExperimentData, then proceeds to prompt for the outcome metric and a list of exogeneous covariates.

\begin{verbatim}
model = experimental_data.add_metric(
  y = ...,
  exogeneous_covariates = ...
).fit(
  LinearModel
)
\end{verbatim}
The LinearModel type trains a model based on the generic form of 2SLS. It passes the instrument, treatment variable, clustering variable, and propensity to create the right form for the model, although the various forms are merely special cases of a a single numeric solver. This simple and short API captures complexity across different types of experiments and different types of models, and collapses multiplicity into a few lines of code.

\subsection{Heterogeneity}
Creating software for an AI agent to estimate an average effect helps the business to measure returns on their changes. To provide greater insights, we also want the software to estimate conditional average effects, and time dynamic effects. These help businesses understand segments of shoppers, their different behaviors, and ultimately whether product changes should be designed for specific segments. Time dynamic effects inform the business whether returns are accelerating or decelerating, which informs the amount of investment.

Exogeneous covariates are already specified in order to decrease variance; estimating the conditional average effect is a difference in conditional expectations: $\tau(x) = E[y | T = 1, X = x] - E[y | T = 0, X = x]$. Similarly, to estimate the effect over time we want to include a time variable, $t$, in the list of exogeneous covariates and trace the process $\tau(x, t)$ over time. Along the way, we will need to change the granularity of the data from a dataset of units to a repeated observations dataset measured over time. In doing so, we introduce an additional layer of clustering due to repeated observations. However, our framework will negate this complexity by simply asking the user to specify clusters under ExperimentData. 
The set of average effects, conditional effects, time dynamic effects and the corresponding requirement of clustering, increase complexity and multiplicity. Our framework makes querying the effect simple by allowing the user to compose any segment after fitting a model, such as 
\begin{verbatim}
effects = model.infer_effect(
  condition_on = [
    ...
  ]
)
\end{verbatim}
The \texttt{condition\_on} parameter allows the user to provide a boolean expression that operates on the covariates $x$. For example, in our motivating example we may condition on shoppers who are using a desktop computer to visit the online store.

The estimation of the effect is a function of $\beta_1$, $\beta_3$, $X$ and $\varepsilon$ and is not easily derived from standard regression tables. The form of the model informs our compute strategy, which adds another layer of complexity. The complexity is controlled by Delta Vectors \cite{wong2024delta}, the numerical computing strategy that abstracts the form of the model and allows a single path to compute average effects, conditional effects, and time dynamic effects. Since anytime-valid inference is a modification to the standard errors of the linear models, peeking corrected inference can also be reported using Delta Vectors. Therefore, we tame multiplicity at the inference step across all experiment types, models, and effect types.

\subsection{Decisions through Policy Making}

Businesses want to cater their products to their audiences. In our motivating example, the average shopper does not benefit from the new change to diverse products, but some do. Say some holiday shoppers, and shoppers browsing on large desktops, benefit from the treatment. By measuring conditional average effects, we can determine an algorithmic policy that optimizes which shoppers should receive treatment. Given a shopper's feature vector, $x_i$, we can offer them one of many interventions in the action space $\{A_0, A_1, A_2, ...\}$. A reward is computed, which is the conditional average effect, $R(a_i, x_i)$ when choosing action $a_i$ for a user with vector $x_i$. A policy function would be the optimization function that chooses an intervention to maximize the reward, such as. 

\begin{align}
    R(a_i, x_i) &= E[y_i | A = a_i, X = x_i] - E[y_i | A = A_0, X = x_i] \\
    \pi(x_i) &= \arg max_a R(a, x_i)
\end{align}
Many variations of policy functions exist, but for simplicity we focus on Thompson Sampling \cite{Thompson}. This algorithm operates on a posterior distribution of the reward, which will capture both the expectation of the treatment effect and its uncertainty. It can be fit by exchanging the 2SLS model for a bayesian linear regression. After fitting parameters to that regression, we infer the posterior distribution of the effect, just as we did for 2SLS frequentist effects. Using the effects and their uncertainty, we probabilistically rank each action in the action space, ultimately returning probabilities $p(a, x_i)$ that action $a$ is the best action for the shopper. In aggregate, the optimal decision is then to randomly choose an action according to $p(a, x_i)$. To capture this in software, we leverage the preceding section that estimates the effects and its distribution, and extend the API with

\begin{verbatim}
model = experimental_data.add_metric(
  y = ...,
  exogeneous_covariates = ...
).fit(
  BayesianLinearRegression,
  prior = ...
)
effects = model.infer_effect(
  condition_on = [
    ...
  ]
)
effects_rank = effects.infer_rank()
\end{verbatim}
By leveraging the \textit{infer\_effect} function, our framework makes it simple to probabilistically rank the arms in both a noncontextual and contextual bandit, while also gaining variance reduction through exogeneous covariates. For online serving, we can serialize the model and infer the reward in an online environment, then invoke the policy algorithm to decide an action based on the action probabilities.

\begin{verbatim}
reward_distribution = model.infer_reward(
  ContextVector(...)
)
action = Policy(reward_distribution).decide()
\end{verbatim}

Algorithmic policy making allows the business to quickly deploy strategies that personalize their product, and is a way to close an iteration loop. Our framework makes it easy to orchestrate effect estimation with algorithmic decision making, achieving objective \ref{goal4}.

\subsection{Mediation and Attribution}

In the previous subsections we described how to report an estimated effect size. However, effect sizes are not sufficient to reveal what the business should do next. In this subsection we describe how to structure the software framework so that an AI agent can provide insights to understand why the metrics are changing, and help the business focus their development for the next iteration.

In our motivating example, the business deployed a change that resulted in negative returns. Many theories explain why. The first theory discusses the challenge in finding the target item that was searched for. If this is the root cause for the negative returns, then there is at least one plausible improvement; the target item can be displayed with a larger image, while other items that are displayed for diversity are shown with smaller images. Understanding the mechanism, or the pathway for how the change effects the metrics, is crucial for business iteration.

We start by positing a graph that illustrates the paths from treatment to sales. For simplicity, we use a standard conversion funnel below.

\begin{verbatim}
Graph:
    T -> Searches
    T -> Product Views
    Searches -> Product Views ->  Sales
\end{verbatim}
The total causal effect is the effect of the treatment on sales. Along the way, the business needs to know the attribution: is the treatment generating more interest in the store's inventory, and those additional product views are generating more sales? Is the treatment generating more relevant search results, and the reduction in search time is causing an increase in sales? Depending on the attribution, the store may focus on marketing their inventory, or it may focus on their technology stack.
To achieve this insight we can decompose the total causal effect into mediated effects \cite{imai2010identification}: the effect of S on Sales and the effect of V on Sales, where S is searches and V is views. In our mediation analysis, there are many simplifications we can leverage due to having a controlled randomized experiment. At the same time there is complexity in this environment due to having multiple mediators that are related to each other in a funnel. Our framework will leverage the reductions and control the complexity, which once again center on a system of equations as in \cite{imai2010identification, lemardelet2022illustrations}.

\begin{align}
    S &= T \alpha_1 + X \gamma_1 + \varepsilon_1 \\
    V &= T \alpha_2 + S \theta + X \gamma_2 + \varepsilon_2 \\
    Y &= T c + S \beta_1 + V \beta_2 + X \gamma_3 + \varepsilon_3
\end{align}
The direct effect of $T$ on $Y$ is captured by $c$. The indirect effects of $S$ on $Y$, and $V$ on $Y$, are captured by $\alpha_1 \beta_1$ and $\alpha_2 \beta_2$ respectively. The serial path, sometimes also called the waterfall, is captured by $\alpha_1 \theta \beta_2$. Now, we leverage identification properties from the experiment. The parameter $\alpha_1$ is identified due to controlled randomization on $T$, regardless of our choice of $X$. $\alpha_2$ is conditionally identified when $X$ controls for confounders between $T$ and $S$. Our ability to measure the mediated effects reduces to choosing $X$ such that we can reasonably assume unconfoundedness between $T$ and $S$, $T$ and $V$, and $S$ and $V$, while not creating collider bias. Previously, we discussed using $X$ as a set of control variables to reduce the variance of the treatment effect. Those variables should be reused here; since $X$ has the property of being measured prior to treatment, it guarantees that there is no collider bias. $X$ can be safely reused in the software.

To provide a composable framework for objective \ref{goal5}, we start with the total effect, assume a graph, then prompt the user for a mediator. The following code would return the mediated effect, estimated from the serial regressions and reusing the $X$ variable that was used to fit the total effect.

\begin{verbatim}
mediated_effect = effect.with_graph(
  graph = ...,
).mediate_effect(
  mediator = ...
)
\end{verbatim}

\subsection{Projecting Effects}

Results from experiments may not extrapolate well to future economic and political environments. To help businesses adapt, we need to be prepared to inform the business whether their changes will continue to yield positive returns in the future.

Suppose that macroeconomic indicators are being tracked with our example shopping data. We notice that sales are high when economic indicators are strong. Say that the experiment is run during a period of strong indicators, and we report a treatment effect under such an environment. However, we anticipate that indicators will turn and the economy will weaken. We wish to project the treatment effect onto this anticipated environment. To do this, we can measure two conditional treatment effects, conditioning on strong indicators and weak indicators. Then, we can reweight these conditional effects by the probability that we believe they will happen. This will reconstitute a new, projected effect. To do this in our framework we operate on \textit{infer\_effects}, then invoke \textit{project}, achieving objective \ref{goal6}.

\begin{verbatim}
  effects = model.infer_effects(
    condition_on = [
      economic_indicator == 'strong',
      economic_indicator == 'weak'
    ]
  )
  effects.project({
    "economic_indicator == 'strong'": 0.2,
    "economic_indicator == 'weak'": 0.8,
  })
\end{verbatim}

\section{Compute Strategy and Dispatch}

We make our software efficient at detecting effects, and efficient at computation. Our framework exposes five key modules: metric computation, model training, effect estimation, policy training, and mediation analysis. Throughout the stack, we make use of batch processing, a dispatcher for parallel processing, and optimizations for numerical computation.

Before modeling begins, we retrieve data from a warehouse, a step in \textit{add\_metrics}. Our framework is aware of the aggregation and source table associated with each metric. It divides metrics computation into data units, then batch processes multiple metrics that come from a common source table. Metrics that come from unique tables, or from unique units of aggregation, are dispatched across parallel jobs.

After metrics computation, model training is optimized through sparse linear algebra and compression, as in \cite{wong2019efficientcomputationlinearmodel, wong2021you}. Per data unit, we train a single model on multiple metrics simultaneously. In seemingly unrelated regression, the metrics can be horizontally concatenated as a matrix, and solvers for OLS can be reused to fit the model in a single shot algorithm. After training the model parameters, delta vectors \cite{wong2024delta} can be used to efficiently infer multiple conditional effects for multiple metrics. This is a vectorization strategy that makes computation efficient. Not all models support this strategy, so a similar dispatcher creates statistical units that assign a model to a metric and dispatches parallel jobs.

\section{Benchmarks and Discussion}

To demonstrate the value of this software framework, we ask AI to analyze experiment data and develop its own solution to objectives 1 - 6. In contrast, we benchmark that solution with one composed from our framework. We evaluate for correctness, code complexity, and numerical efficiency.

The AI developed a script with 1000 lines of code. It spent 25 minutes to develop the solution, presumably scanning multiple pages of documentations, literature, and APIs. On the simple task of computing an average treatment effect, the solution it provided was mathematically correct, as expected. However, it did not initially produce the correct solution for the harder objective of estimating average treatment effects when the treated variable, $T$, contains selection bias. In this scenario, it uses the instrument to compute the local average effect, but fails to remap that to the average effect. Upon careful and precise prompting, we do acknowledge that it can achieve a correct solution. It also fails in using an instrument on clustered data. While it was able to defer to \textit{statsmodels} for the average effect, it attempted to implement its own algebraic solution for the case of two stage least squares with clustered data. In this case, we presume that the AI had to not only scan API documentation, it had to reference econometrics literature to obtain the algebraic solution, then implement it using \textit{numpy}. Its implementation was ultimately incorrect as it failed to use the projection matrix of the first stage correctly. Finally, the AI was not able to implement the mediation analysis correctly, where it incorrectly assumed parallel mediators instead of serial mediators. We claim the AI achieves 3 out of the 6 objectives stated in this paper.

Beyond correctness, the AI struggled to produce solutions with consistency. Despite being prompted to use variance reduction, only its solution for average effects made use of the covariates, $X$. Its solution to 2SLS did not utilize the covariates, nor was it used for developing a contextual multiarm bandit for policy algorithms. The combination of incorrectness and inconsistency demonstrates the challenges in multiplicity and orchestration. A simple change to the data context, like clustered data, has multiple consequences to how models are fit and how effects are inferred. In addition, orchestrating a consistent solution across multiple disjoint strategies, such as variance reduction, instrumental variables, and policy making, is hard without a grammar.

Our framework provides a simpler solution using 120 lines of code. The AI agent equipped with our framework correctly achieves all objectives. In the framework we implemented a model reduction strategy that frames all linear effects as a special case of 2SLS with clustered data, so it does not face complexity. The AI makes consistent use of covariates all throughout the prompts for anytime valid inference, policy making, and mediation.

Other than code complexity and code correctness, the solution provided by the AI took 25 minutes on a dataset of $n = 1e6$ observations. As it progressed through the objectives, it required 6 GB of memory to sustain its solution as it orchestrated many technologies to solve the problems, and as it created multiple copies of data and matrices to solve for multiple treatments, multiple outcome metrics, and multiple segments. Its compute strategy was naive, where it used many nested loops to execute multiplicity. It did not batch compute, vectorize, or use specialized solutions like seemingly unrelated regression. The AI deferred to sampling the input data first in order to successfully execute its analysis. Our solution is memory efficient due to leveraging intermediate solutions and compression effectively. It is also numerically efficient due to being optimized for sparse algebra,  delta vectors, and vectorization to train multiple models for multiple outcome metrics simultaneously. Our solution completes the analysis with 100\% correctness and without sampling in 3 minutes. It consumes 60 MB of memory.

\section{Conclusion}

Businesses need to quickly adapt to new trends, new environments, and new customers. They need to use data to understand the theory of the business, evolve it, and determine what experiments will advance the theory in order to create features that generate positive returns. 
By improving experimentation speed, AI will be able to contribute to ideation and evaluation in a complete loop. Through composable software, we streamline the human-AI interaction and are able to easily extend the evaluation logic as needed.

This paper provided a unique point of view on experimentation: we leverage multiple mathematical reductions across causal inference frameworks in order to create a single software framework, which enables the creation of an AI agent that can analyze online experiments for businesses. A single and comprehensive framework is important to have minimal errors when orchestrating multiple causal analyses, ensure software safety, and be efficient with multiplicity.

Our framework maps various randomization strategies from experimental design to the appropriate statistical models that can estimate the causal effect. The models report the causal effects efficiently, using variance reduction as well as anytime valid techniques to deliver insights to the business faster. In addition, the models can infer different types of effects that are suitable for user segmentation and trend analysis, helping the business to ideate and evaluate opportunities for personalization. Without our reduction strategies, chaining these capabilities may require an AI to navigate over 100 combinations of tools. With our reductions, many combinations of functions are reduced to a single interface. 
In addition to these reductions, our software framework is numerically efficient. This results in benchmarks that show a significant reduction in the number of lines of code to analyze an experiment, the amount of time the AI agent spends, and an increase in correctness.

After reporting on these effects, the user can prompt the agent to search for the mechanism for how the treatment induced a change in the metrics. In our example, this capability from the AI agent helps the business understand whether a shopper is experiencing friction, decision fatigue, or poor performance on the site. That understanding enables targeted business iterations. The agent can also develop policies as a way for the business to deliver the best treatment to each user.

Our contribution furthers the AI loop. AI enables creative thinkers from the business to create new business ideas, do market research, and build a prototype. Experimentation services let businesses execute an experiment. Our AI agent analyzes those experiments with rigorous causal effects analysis, and provides insights for the next iteration, which can be developed with AI again.

\bibliographystyle{ACM-Reference-Format}
\bibliography{references}


\end{document}